\documentclass[11pt]{article}
\usepackage{color}
\usepackage[english]{babel} 
\usepackage{amsmath,amsfonts,amsthm,bm} 
\usepackage{mathrsfs}
\usepackage{amssymb}
\usepackage{amsthm}
\usepackage{epstopdf}
\usepackage{graphicx}
\usepackage{mathtools}
\usepackage[usenames,dvipsnames, table]{xcolor}
\usepackage[hang]{caption}
\usepackage{hyperref}

\usepackage{oldgerm}  
\usepackage[yyyymmdd,hhmmss]{datetime}
\usepackage{dsfont}

\usepackage{mathabx} 
\def\bee{\begin{enumerate}}\def\eee{\end{enumerate}}
\def\bei{\begin{itemize}}\def\eei{\end{itemize}}
\newcommand{\nco}{\newcommand}
\def\R{\mathbb{R}}
\nco{\red}{\color{red}}
\nco{\blue}{\color{blue}}
\nco{\cyan}{\color{cyan}}
\nco{\brown}{\color{Magenta}}

\nco{\magenta}{\color{magenta}}

\nco{\violet}{\color{violet}}
\nco{\olive}{\color{Emerald}}
\nco{\orange}{\color{orange}}
\nco{\redend}{\normalcolor}
\nco{\blueend}{\normalcolor}

\def\ommit#1{{}}
\def\({\left(}
\def\){\right)}
\def\ie{{\it i.e.,\/}\ }
\def\ie{{\rm i.e.,\/}\ }

\nco{\rnc}{\renewcommand}
\rnc{\title}[1]{{\Large\bf\mbox{}\\\medskip#1\bigskip\medskip\\}}
\rnc{\author}[1]{{\large #1\smallskip\\}}
\nco{\address}[1]{{\em #1\medskip\\}}

\def\ii{\mathrm{i\,}}
\nco{\bun}{{\bf 1}}
\def\be{\begin{equation}}\def\ee{\end{equation}}
\def\bea{\begin{eqnarray}}\def\eea{\end{eqnarray}}
\def\bee{\begin{enumerate}}\def\eee{\end{enumerate}}
\def\bei{\begin{itemize}}\def\eei{\end{itemize}}
\def\oh{\frac{1}{2}}
\def\ommit#1{{}}
\def\su{{\rm su}}
\def\SU{{\rm SU}}\def\U{{\rm U}}

\def\eq=#1{\buildrel #1 \over{=}}

     \def\CJ{{\mathcal J}} \def\CK{{\mathcal K}}

\def\c{c}  \def\c{r}    

\def\ii{\mathrm{i\,}}

\def\R{\mathbb{R}}

\def\T{\mathbb{T}}

\begin{document}

\begin{titlepage}
%
\begin{center}
\title{Multiplicities, pictographs, and volumes}
\medskip
\author{Robert Coquereaux} 
\address{Aix Marseille Univ, Universit\'e de Toulon, CNRS, CPT, Marseille, France\\
Centre de Physique Th\'eorique}
\bigskip\medskip
\medskip

\bigskip\bigskip
\begin{abstract}
The present contribution is the written counterpart of a talk given in Yerevan at the SQS'2019 International Workshop ``Supersymmetries and Quantum Symmetries''  (SQS'2019, 26 August - August 31, 2019).
After a short presentation of various pictographs (O-blades, metric honeycombs) that one can use in order to calculate $\SU(n)$ multiplicities (Littlewood-Richardson coefficients, Kostka numbers), we briefly discuss the semi-classical limit of these multiplicities in relation with the Horn and Schur volume functions and with the so-called $R_n$-polynomials that enter the expression of volume functions. For $n \le 6$ the decomposition of the $R_n$-polynomials on Lie group characters is already known, the case $n=7$ is obtained here.

\end{abstract}
\end{center}

 \end{titlepage}
 
 
 \def\x{{\magenta x}}  
\def\x{x}
\section{Introduction}

The first part of this paper contains a brief summary of several relatively new methods that allow one to determine the multiplicity $C_{\lambda, \mu}^{\nu}$ of an irreducible representation (irrep) $\nu$ in the tensor product $\lambda \otimes \mu$ of tho irreps of the Lie group $\SU(n)$, aka  Littlewood-Richardson (LR) coefficients, and also the multiplicity $K_{\lambda, \delta}$ of a weight $\delta$ in the weight system of an irrep characterized by its highest weight $\lambda$ (aka Kostka numbers). These methods (O-blades, metric honeycombs, lianas) are exemplified in the case of $\SU(5)$.

\noindent
The behavior of multiplicities, for very large weights, is determined by the so-called volume function(s).
Those functions, which appear as a kind of semi-classical limit of  usual multiplicities, can be obtained in terms of Lie group characters but their expression also involves a particular multivariate invariant polynomial $R$, depending only on the chosen Lie group, and which, in the case of $\SU(n)$, is explicitly known only for small values of $n$. 
In the second part of this paper, after a short description of several general features of the volume functions and of the $R$-polynomial for $SU(n)$, we illustrate several properties of the latter in the case $n=5$. 
We also give its explicit expression when $n=7$ (expressions for the cases $n \leq 6$ can already be found in sec.~4.2 of \cite{CZ1}).
The appendix contains some other information and explicit results about the structure of those invariant $R$-polynomials for cases other than $\SU(n)$.

\section{Multiplicities}
 \subsection{LR-coefficients for $\SU(n)$ : O-blades and metric honeycombs}
 Here we take $\SU(5)$ as an example but the reader will have no difficulties to generalize the concepts.
 Consider two irreducible representations defined by their highest weights $\lambda=\{3, 4, 3, 5\}$ and $\mu=\{4, 3, 5, 4\}$ in Dynkin coordinates (components on the basis of fundamental weights), and their tensor product denoted $\lambda \otimes \mu$.
 These two irreps have respective dimensions\footnote{The dimension of the irrep $\lambda$ is found from the formula {\scriptsize $$\text{dim}(\lambda) =
\frac{1}{288} (1 + \lambda_1) (1 + \lambda_2)(1 + \lambda_3)(1 + \lambda_4) 
(2 + \lambda_1 + \lambda_2) (2 + \lambda_2 + \lambda_3) (2 + \lambda_3 + \lambda_4) 
 (3 + \lambda_1 + \lambda_2 + \lambda_3) (3 + \lambda_2 + \lambda_3 + \lambda_4) 
 (4 + \lambda_1 + \lambda_2 + \lambda_3 + \lambda_4)$$}}
$5001750$ and $9281250$.
Their tensor product is a reducible representation whose decomposition contains $5223$ distinct irreducible representations, each of them coming with a multiplicity usually larger than $1$: 
the smallest multiplicity that occurs in the tensor product is $1$, the largest is $1657$, the total multiplicity (sum of multiplicities) is $1043606$.
Among the terms of the decomposition, we select the irrep $\nu$ with highest weight $\{2, 2, 4, 2\}$. This irrep has multiplicity $371$ in the tensor product, \ie $C_{\lambda, \mu}^{\nu}=371$.
There are several ways to obtain this last information. 
The drawback of the traditional method using Young diagrams (with at most $4$ lines, since we are in $\SU(5)$) is that one would have to determine all the terms of decomposition, even if we are only interested in the multiplicity of $\nu$.

\begin{figure}[!tbp]
  \centering
\includegraphics[width=12pc]{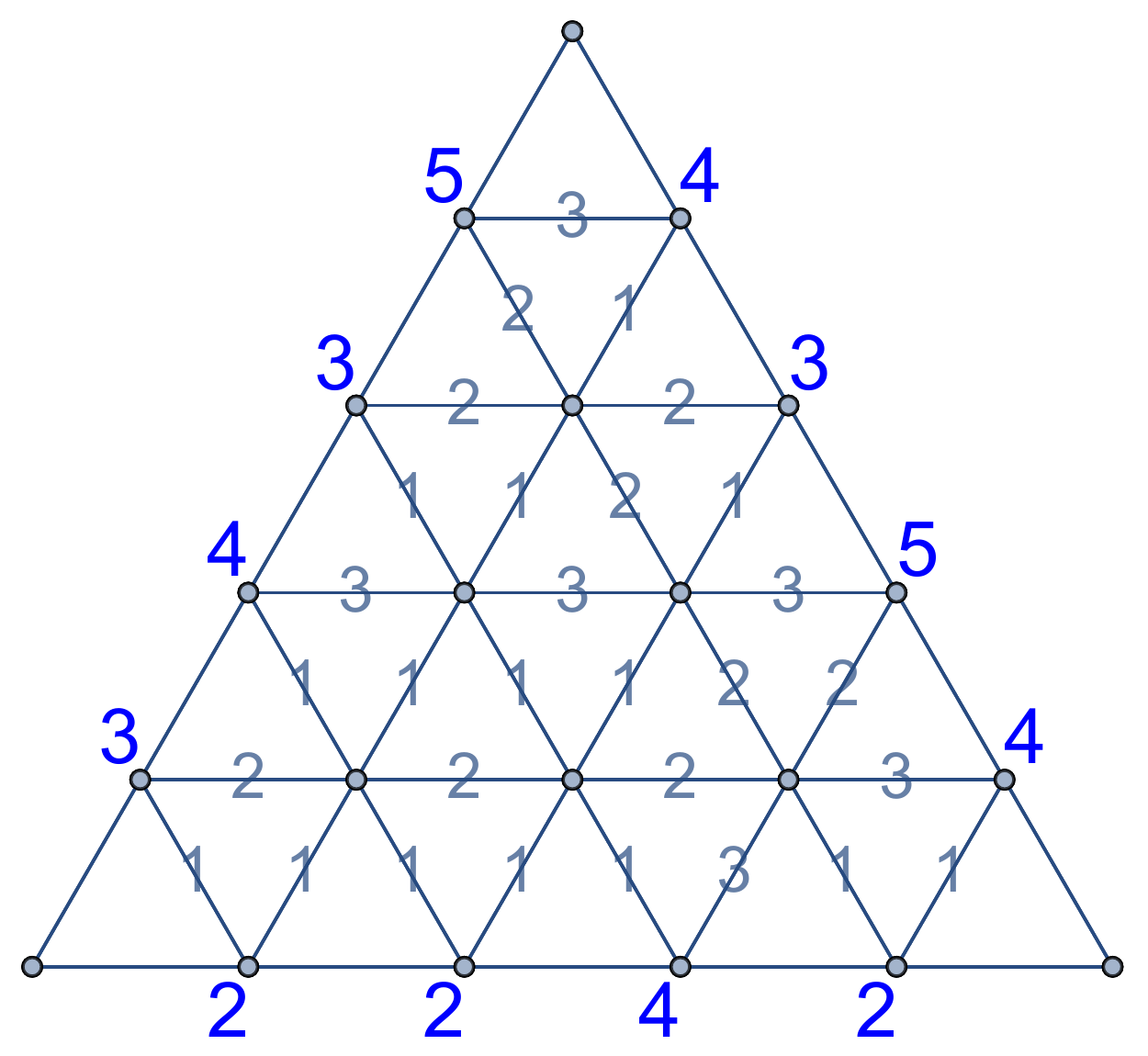}
 \hspace{2.cm}
\includegraphics[width=12pc]{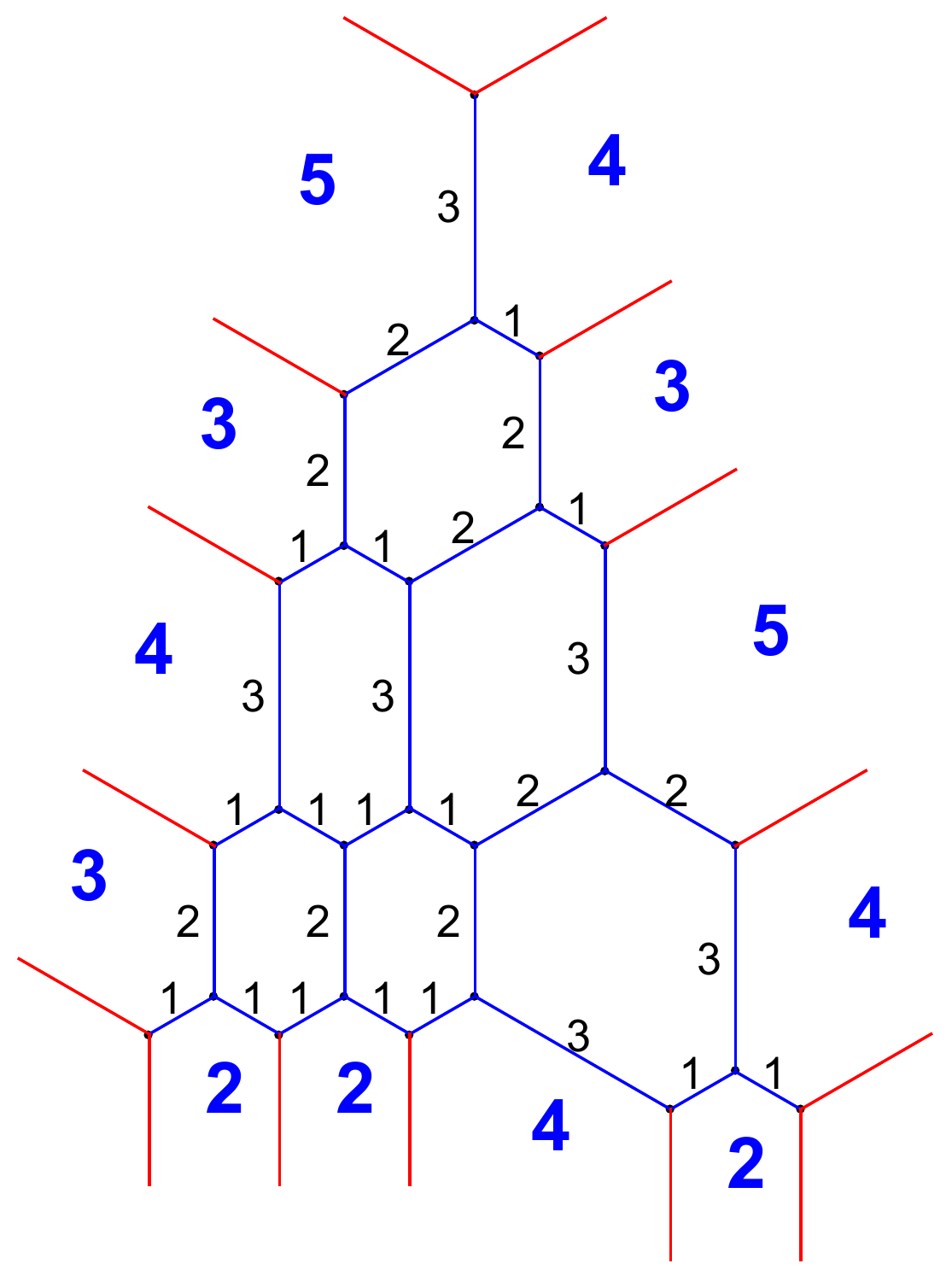}
\caption{   \label{AnSU5ObladeHoneycomb} 
    Left: One of the $371$ O-blades for the $\SU(5)$ tensor product branching rule:\\ $\lambda \otimes \mu \mapsto \nu$ with $\lambda=\{3,4,3,5\}$,  $\mu=\{4,3,5,4\}$, $\nu=\{2,2,4,2\}$.\\
    Right: Its associated metric honeycomb.}
 \end{figure}

Another method uses honeycombs. Honeycombs were invented by \cite{KT00} to solve the Horn problem. The honeycombs described in the latter reference are triangular arrays whose sides are parametrized by partitions (Young diagrams). They can also be used to solve multiplicity problems for unitary groups. 
Since our weights are characterized by Dynkin coordinates, not by partitions, and since we are dealing with $\SU(n)$, not with $U(n)$, it is easier to use a variant of this construction: the {\sl metric honeycombs}, also called {\sl isometric honeycombs}, that have sides parametrized by the Dynkin coordinates of weights.
Actually, it is even simpler to start with a kind of dual of this construction, the so-called {\sl O-blades}.  The author learned them from \cite{Ocn}, about ten years ago, they have been described in one of the last sections of \cite{CZ3}.
Let us consider the following picture, Fig~\ref{AnSU5ObladeHoneycomb} (left).

The rules are simple and are as follows. \\
One first labels the three sides of the triangles with the Dynkin components of $\lambda$, $\mu$ and $\nu$.
Notice that the ordering of components is clockwise for  $\lambda$ and $\mu$ but counterclockwise for $\nu$ (one could choose the clockwise direction for all three weights, and this would seem more ``symmetrical'' but then one would be looking at the branching rule $\lambda \otimes \mu \otimes \overline{\nu} \mapsto \mathds{1}$, which has the same multiplicity).
The game is then to complete the picture and decorate the inner edges with non-negative integers (so $0$ is allowed), by respecting the following constraints.\\
There is a conservation rule at each external vertex, for instance $5$ (on the right edge) gives $3+2$.\\
Around each internal vertex ($6$ of them in the $\SU(5)$ case), we have six angles, the ``value of an angle'' is defined as the sum of its adjacent edges, and the constraint is that``opposite angles'' should be equal.
For instance, in the same picture, around the last vertex of the first floor, we have the three equalities $3+1=2+2, 1+3=2+2, 3+2=2+3$.

\begin{figure}[!tbp]
  \centering
     \includegraphics[width=30pc]{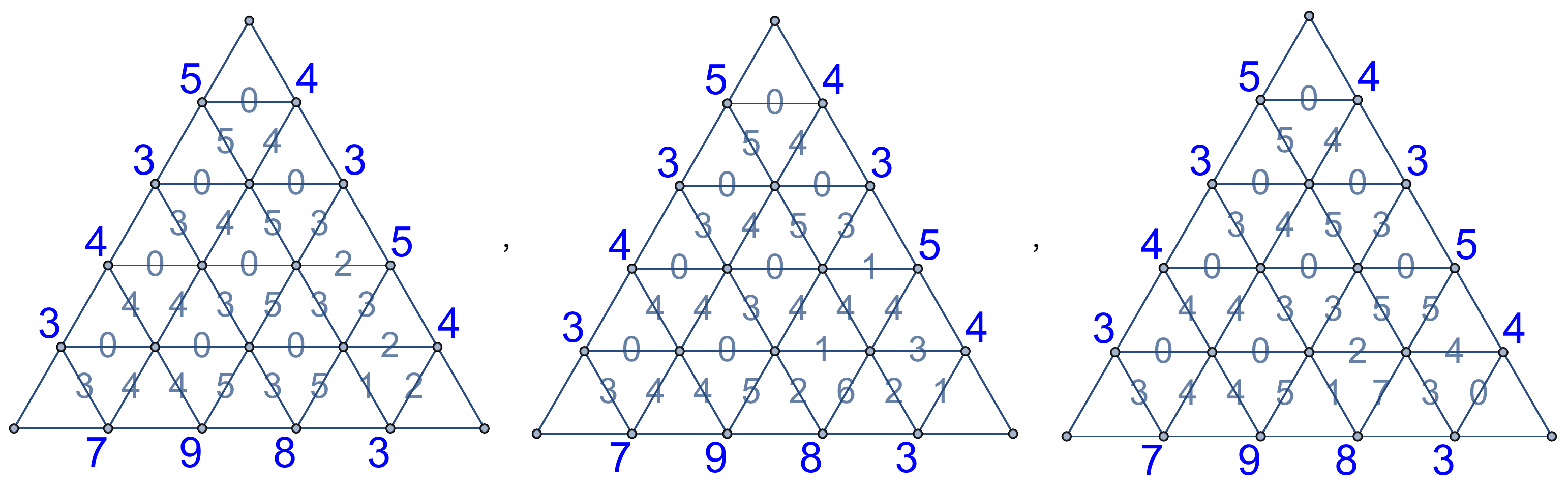}
    \caption{\label{SU5ObladesMultiplicity3}  The three O-blades for the $\SU(5)$ tensor product branching rule: $\lambda \otimes \mu \mapsto \nu$ with $\lambda=\{3,4,3,5\}$,  $\mu=\{4,3,5,4\}$, $\nu=\{7,9,8,3\}$.}
\end{figure}

For the given weights $\lambda$, $\mu$, $\nu$, there are $371$ solutions, \ie $371$ such O-blades, and this is the sought-after multiplicity.
Of course we shall not display all of them, but writing a computer program for this problem is easy.
Just to show one example where we can display all the solutions in a single page, we choose the same $\lambda$ and $\mu$ as before, but take $\nu =\{7,9,8,3\}$. 
The reader can convince himself (herself) that there are only three solutions, they are displayed in Fig~\ref{SU5ObladesMultiplicity3};  the multiplicity in this case is therefore equal to $3$.

Notice that in this last example, many inner edges have value $0$. This was not the case for the O-blade chosen in Fig~\ref{AnSU5ObladeHoneycomb}(left), which, for this reason, may be called a ``non-degenerated'' O-blade\footnote{There is actually only one non-degenerated O-blade among the $371$ solutions, this is the one displayed in Fig~\ref{AnSU5ObladeHoneycomb}.}.

Under duality (take lines orthogonal to the inner edges of the O-blade), inner vertices become parallelo-hexagons, and the O-blade itself becomes an isometric honeycomb. See Fig.~\ref{AnSU5ObladeHoneycomb}(right).

\begin{figure}[!tbp]
  \centering
     \includegraphics[width=12pc]{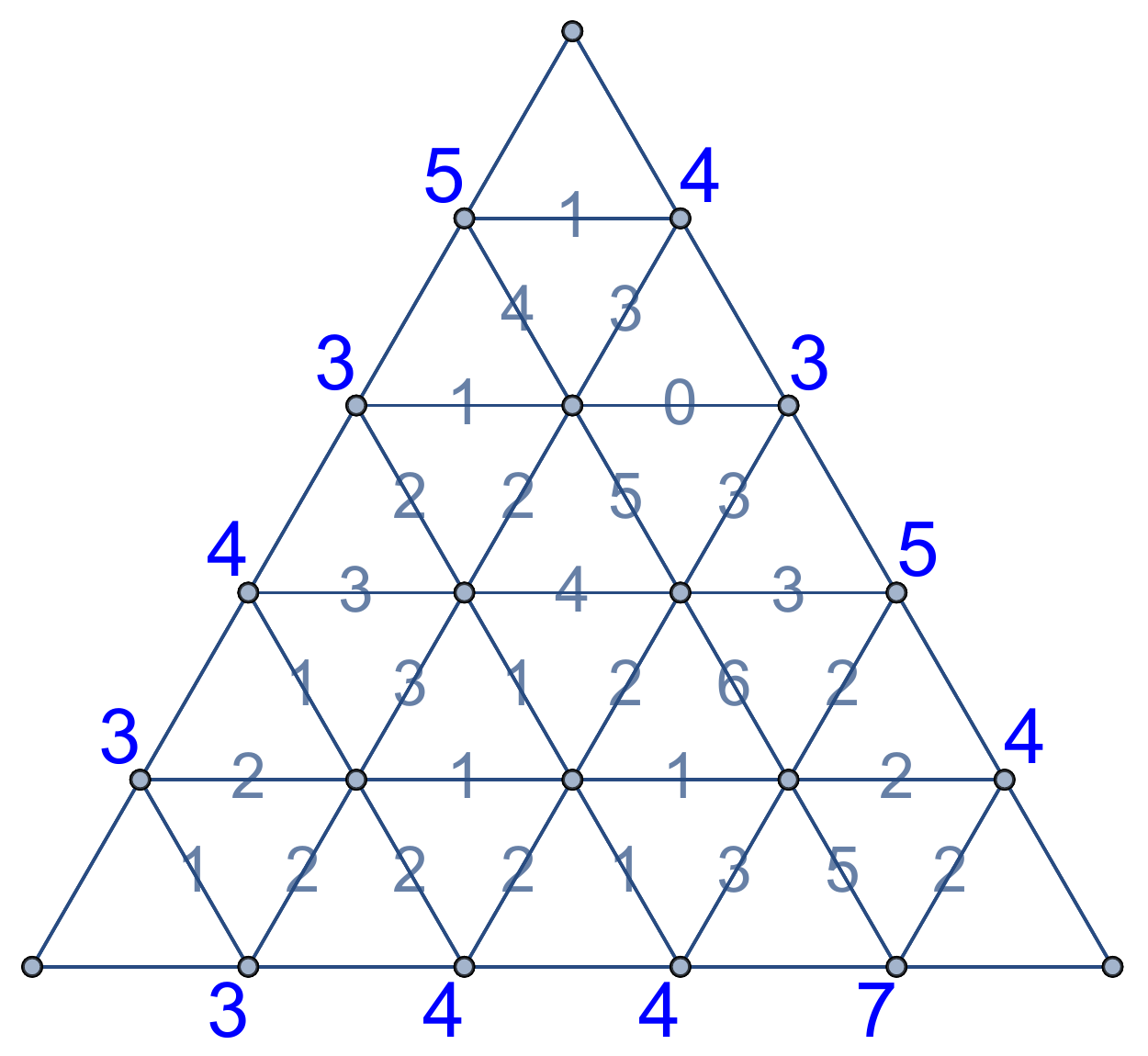}
     \hspace{2.cm}
          \includegraphics[width=12pc]{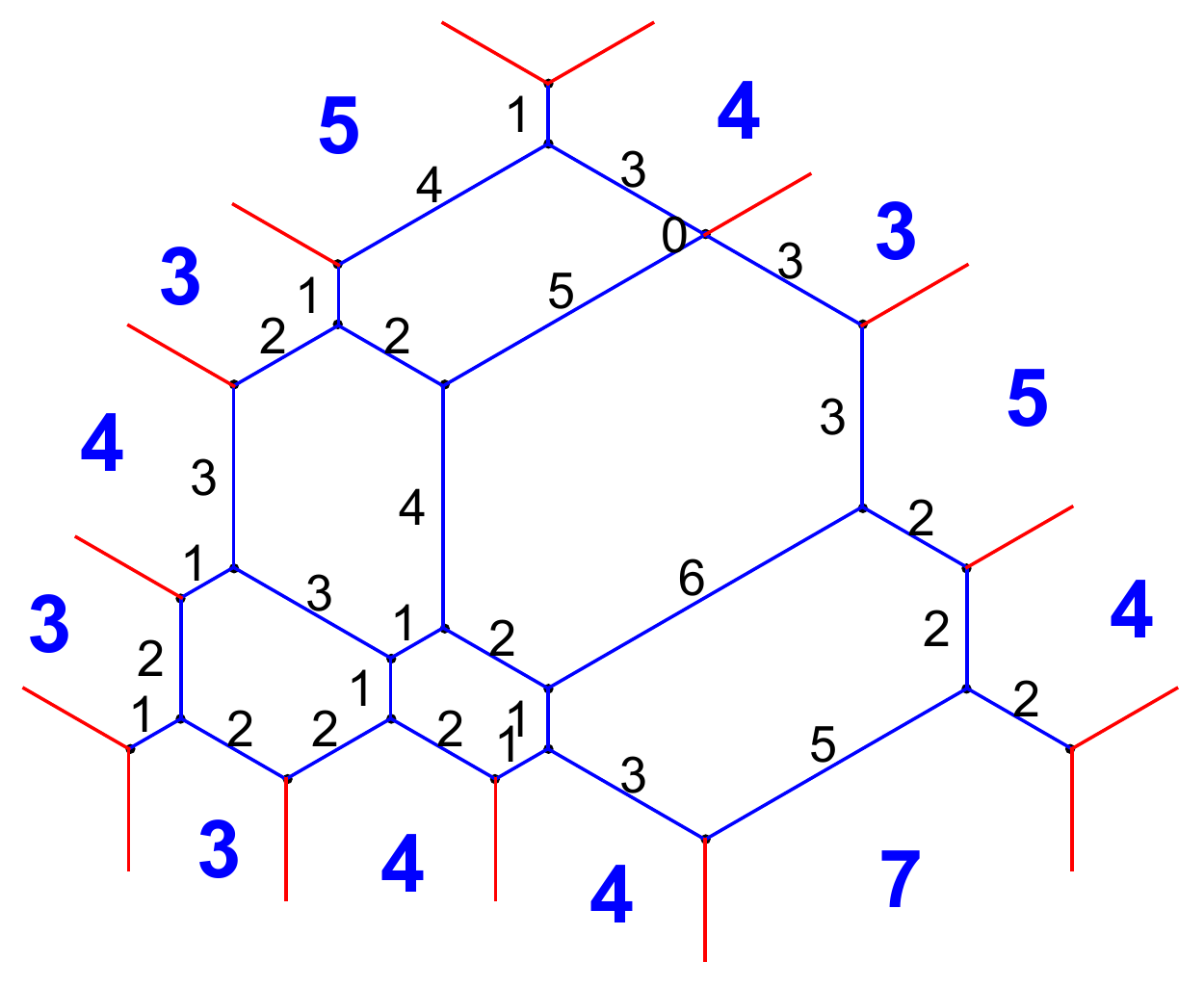}
    \caption{\label{AnotherSU5pictograph}  One of the $1427$ O-blades and its associated metric honeycomb for the $\SU(5)$ tensor product branching rule: $\lambda \otimes \mu \mapsto \nu$ with $\lambda=\{3,4,3,5\}$,  $\mu=\{4,3,5,4\}$, $\nu=\{3,4,4,7\}$.}
\end{figure}

The inner constraint now reads : sums of opposite pairs of opposite edges of hexagons are equal. 
This is why the hexagons that appear in these honeycombs are parallelo-hexagons and it explain why these honeycombs are ``metric'' : the labels carried by edges are indeed the lengths of these edges.
Notice that the particular honeycomb drawn in Fig.~\ref{AnSU5ObladeHoneycomb}(right) is not degenerated because all its hexagons are genuine hexagons; for other examples, and since edges can in general vanish, the hexagons can degenerate to pentagons, quadrilaterals, etc. 
Another example of an $\SU(5)$  O-blade and of its associated honeycomb is given on Fig.~\ref{AnotherSU5pictograph}; this one is degenerated: one of the six hexagons is actually a pentagon.
 
Drawing metric honeycombs while respecting their metric structure, can lead to an awkward picture in the presence of many edges of vanishing length,  it is then clearer to draw the corresponding O-blades. 
The drawing of metric honeycombs provides however a nice geometrical interpretation of what multiplicity is:  one can go from one honeycomb to another by inflating  or deflating hexagons. In other words, for fixed external sides (determined by the Dynkin components of the three highest weights),  hexagons can breathe, still obeying the integrality constraints, and this freedom gives rise to non-trivial multiplicity.

Honeycombs and O-blades are both simply related to the older Berenstein-Zelevinsky triangles \cite{BZ}, that we shall not describe here; they are discussed for example in the textbook \cite{DFMS}.

\bigskip

Before ending this short presentation we should mention the following: 

$\bullet$ O-blades for $\SU(n)$ have $(n-1)(n-2)/2$ inner vertices (obvious), and, of course, honeycombs contain the same number of (possibly degenerated) hexagons.

$\bullet$ The $3 \times n(n-1)/2$ non-negative  integers labeling the inner edges of the $\SU(n)$ O-blades, in the three directions, are components of the three Kostant vectors $\lambda + \mu - \nu$,  $\lambda + \overline{\nu} - \overline{\mu}$, 
$\mu + \overline{\nu} - \overline{\lambda}$, that belong to the root lattice, on the over-complete family of positive roots (this remark provides a rationale for the existence of these various kinds of pictographs (in the case of BZ-triangles, see for instance \cite{BZ1} or \cite{DFMS}).

$\bullet$ One may consider any O-blade as a linear combination (with non-negative coefficients) of characteristic functions associated with the set of its inner edges.  As such, O-blades can be added, and it is easy to see that they are not independent (existence of syzygies): there is one relation at each inner vertex (see section 4.2.2 of \cite{CZ3}  for a detailed discussion of the $\SU(3)$ case). 
 One can define ``elementary pictographs'' (for instance elementary O-blades):  $3 (n-1)$ of them describe branching rules of the type $f \otimes \overline{f} \mapsto \mathds{1}$, $\mathds{1} \otimes f \mapsto f$ or $f \otimes \mathds{1} \mapsto f$, where $f$ is a fundamental irreducible representation, and $2\times (n-1)(n-2)/2$ describe branching rules of the type $f_1 \otimes f_2 \mapsto f_3$ where the $f_j$'s are fundamental and such that the obtained multiplicities are not equal to $0$. Taking syzygies into account 
 one can show that arbitrary O-blades depend on  $3 (n-1) + 2\times (n-1)(n-2)/2 - (n-1)(n-2)/2 = (n+4)(n-1)/2$ parameters, some of them possibly negative because of the syzygies.  For fixed external sides (fixed values of the highest weights $\lambda, \mu, \nu$), the number of parameters goes down to $d=(n+4)(n-1)/2 - 3(n-1) = (n-1)(n-2)/2$, which is also the number of inner vertices.

$\bullet$ Filling an O-blade (or a honeycomb) with prescribed values of the three highest weights involves a set of constraints that define a family of hyperplanes in $\R^{d}$, ($d=6$ in the $\SU(5)$ case),  but the condition that edges should be non-negative implies that solutions belong only to one of the two half-spaces defined by each of those hyperplanes. Moreover edges should be integral, hence the problem to be solved amounts to find the number of integral points in a polytope defined as an intersection of hyperplanes with integer coefficients. 
Each solution (each O-blade, each honeycomb) is an integral point of a polytope called the {\sl hive polytope}\footnote{Hive polytopes are closely related to BZ-polytopes (see \cite{CMSZ}) but one can identify them for the purpose of counting arguments; they should not be confused with the Horn polytope (or tensor polytope) whose integer points are the distinct irreps $\nu$ in the tensor product $\lambda \otimes \mu$ and  live in a space of dimension $n-1$,  or with the weight polytopes, which are also generically of dimension $n-1$ and are defined as the convex hull of the Weyl orbit of a weight.}(the word ``hive'' alone may have another meaning in the literature \cite{KT99}).\\
Warning-1: such a polytope is not necessarily the convex hull of its integral points since the intersection of several hyperplanes with integer coefficients can sometimes be a vertex with rational, but non necessarily integer components. 
 In other words, those polytopes are rational but not necessarily integral polytopes -- this however occurs only when $n\geq 5$,  see examples in \cite{CZ1}. \\
Warning-2:  hive polytopes are generically of dimension $d$ (given before) but their dimension can be smaller. This occurs when one, or more than one, of the highest weights defining it lies on one of the walls of the dominant Weyl chamber (presence of zeroes in its Dynkin components).
 
 \subsection{Kostka numbers for $\SU(n)$: reduced O-blades and lianas forests}
 The two problems LR-coefficients versus Kostka numbers, are not independent; indeed, for integral weights, one can prove that   
 \be K_{\lambda, \delta} = \lim_{p \to  \infty} \, C_{\lambda, p\rho}^{\delta + p\rho} \label{LRversusKostka}\ee 
 where $\rho$ is the Weyl vector.
 The convergence is rather fast and one can find a bound $b_{\lambda, \delta}$ such that, for $p > b_{\lambda, \delta}$, both expressions  $K_{\lambda, \delta}$ and $C_{\lambda, p\rho}^{\delta + p\rho}$ become equal. 
 One can also find a bound\footnote{A value of this bound was obtained in \cite{CZ:Kostka}, it was then improved and generalized for all semi-simple Lie algebras by \cite{Jeralds:bound}. For $\SU(n)$ it reads $b_{\lambda}=\sum_j{\lambda_j}$ where $\lambda_j$ are the Dynkin components of the weight $\lambda$.} $b_{\lambda}$ such that, for $p > b_{\lambda}$, the equality holds for all the weights $\delta$ of the weight system of $\lambda$. 
 
 In practice, it is simple and instructive to calculate $C_{\lambda, p\rho}^{\delta + p\rho}$ for consecutive values of $p$. Reverting to the $\SU(5)$ example considered in the first part, $\lambda= \{3, 4, 3, 5\}$, taking for instance $\delta = \{0, 3, 0, 1\}$ and using $\rho = \{1,1,1,1\}$, one obtains the following sequence, for $p=1\ldots 10$, which stabilizes:    \[C_{\lambda, p\rho}^{\delta + p\rho}= 0, 103, 685, 1198, 1424, 1468, 1473, 1473, 1473, 1473,\ldots \]
The multiplicity of $\delta$ in the weight system of $\lambda$ is $K_{\lambda, \delta}=1473$.

Now that we have pictographs such as O-blades and metric honeycombs at our disposal, we can use them to compute and display the results (it is better here to use O-blades: drawing the corresponding honeycombs in a metric way is usually not convenient, some of the hexagon edges becoming rather large when the scaling parameter $p$ grows).
One of the three directions of the O-blades becomes actually irrelevant when the scaling parameter $p$ grows, because of stationarity of the previous sequence and since $p$ does not enter the definition of the Kostka number $K_{\lambda, \delta}$ anyway.

\begin{figure}[!tbp]
  \centering
     \includegraphics[width=12pc]{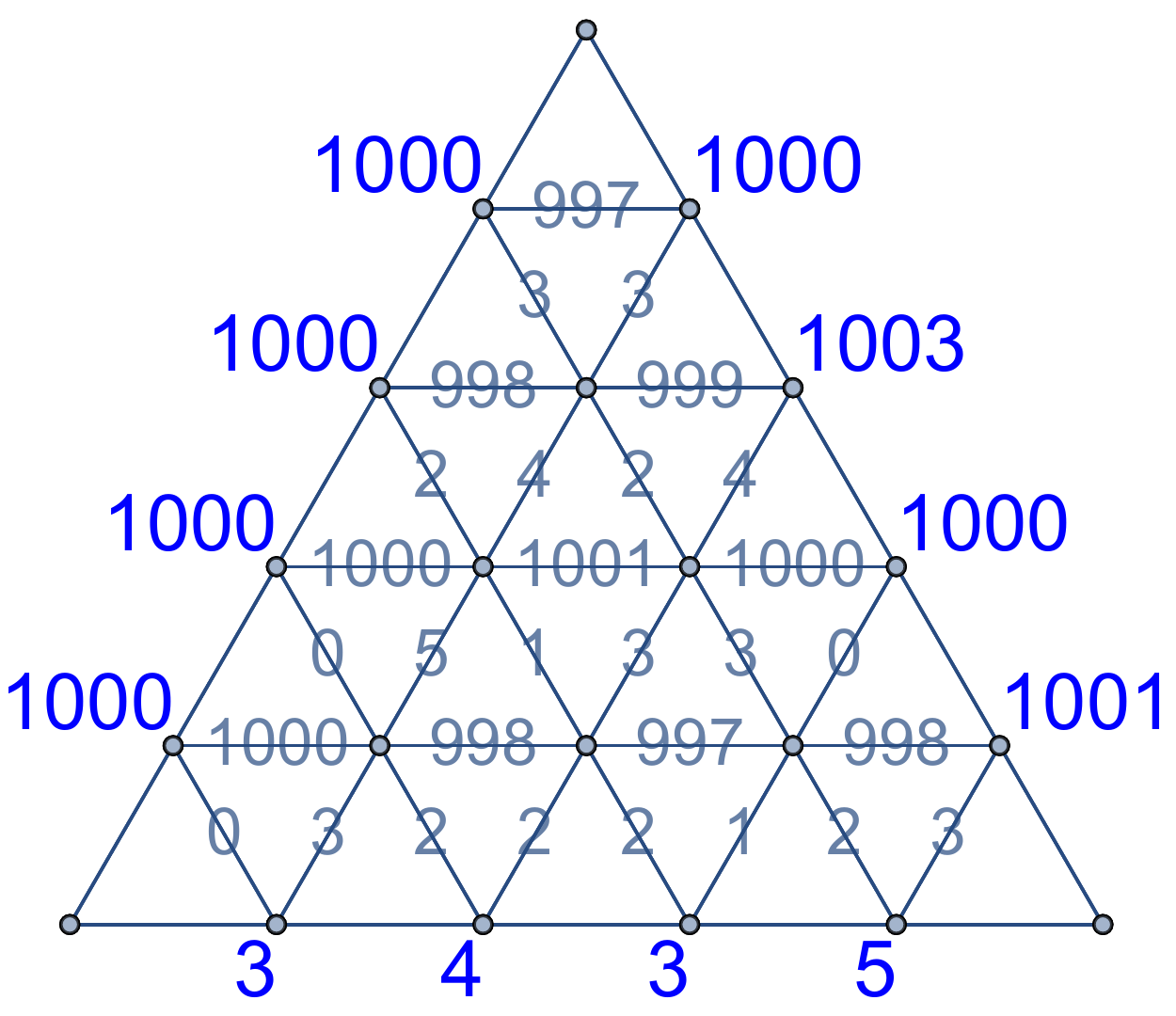}
        \hspace{2.cm}
      \includegraphics[width=12pc]{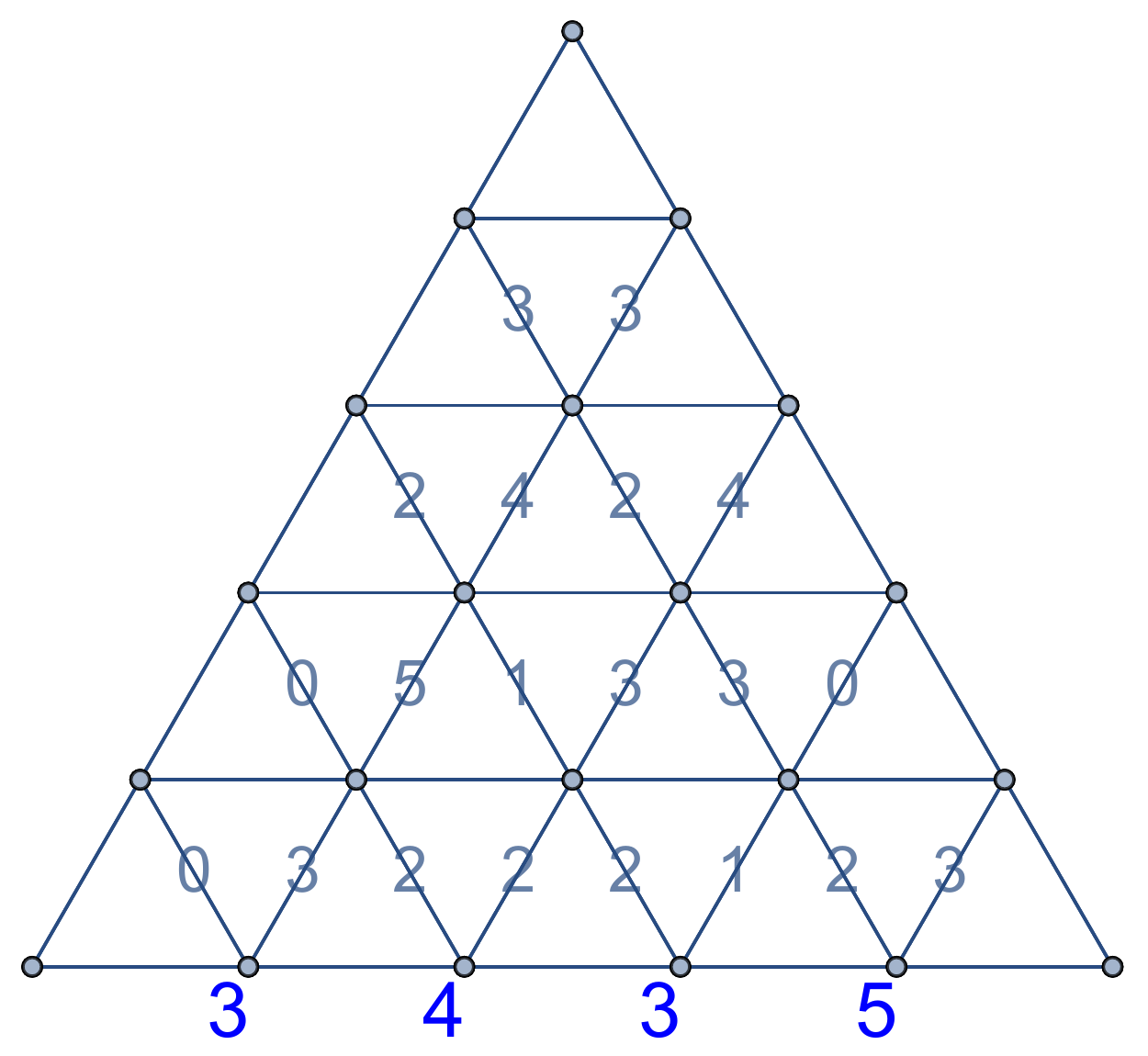}
    \caption{\label{fullobladeforkostka}  
    Left: One of the $1473$ O-blades for the $\SU(5)$ tensor product branching rule: \\$p\rho \otimes (\delta + p\rho)  \mapsto \lambda$ with $\lambda=\{3,4,3,5\}$,  $\delta=\{0,3,0,1\}$ and $p=1000$. \\
    Right: Same, where we removed labels from left and right sides and from horizontal lines.}
\end{figure}

This is most easily seen by looking at the following example, Fig.~\ref{fullobladeforkostka}, where we have chosen $p=1000$;  in order to display the weight $\lambda$ on the bottom, we choose to permute the arguments and draw an O-blade associated with the branching rule $p\rho \otimes (\delta + p\rho)  \mapsto \lambda$ since $C_{p\rho, \delta + p\rho}^{\lambda}=C_{\lambda, p\rho}^{\delta + p\rho}$.\\ The figure on the right is an example of what was called a {\sl reduced O-blade} in \cite{CZ:Kostka}.

A traditional method for determining the multiplicity of a weight in a weight system is to draw all semi-standard Young tableaux for a given partition (Young diagram) and a specified weight. One can also use Gelfand-Tsetlin patterns.
There is another way, using {\sl lianas forests}: Lianas are zigzag lines going upward from an horizontal edge (marked with the Dynkin components of a chosen highest weight $\lambda$) and obeying a number of constraints (specified by $\delta$).
Superposition of lianas builds a liana forest.
The number of liana forests (for fixed $\lambda$ and $\delta$)  is equal to the dimension of the weight subspace defined by $\delta$ in the representation space of highest weight $\lambda$, \ie the multiplicity $K(\lambda, \delta)$.
Intuitively lianas forests replace the semi-standard Young tableaux. One of their interest lies in the fact that each liana forest can be graphically identified with a specific reduced O-blade, so that one does not have to use LR-coefficients (or full O-blades) and formula \ref{LRversusKostka} as an intermediate step.

\begin{figure}[!tbp]
  \centering
     \includegraphics[scale=0.25]{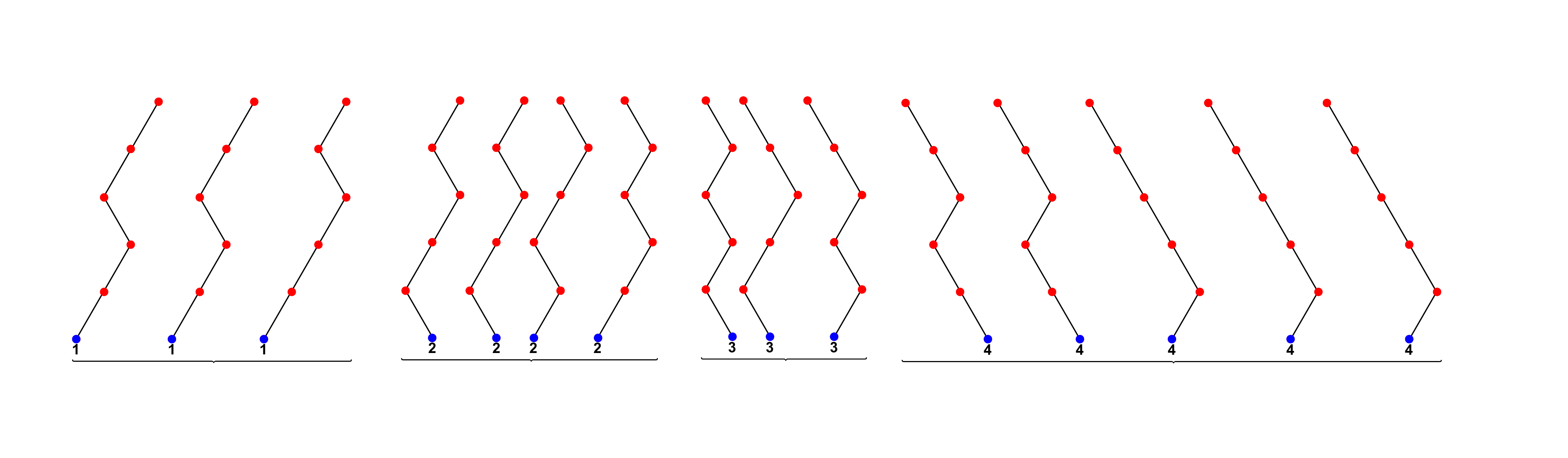}
        \hspace{0.8 cm}
      \includegraphics[scale=0.22 , trim=5cm -2.3cm 0 0 ]{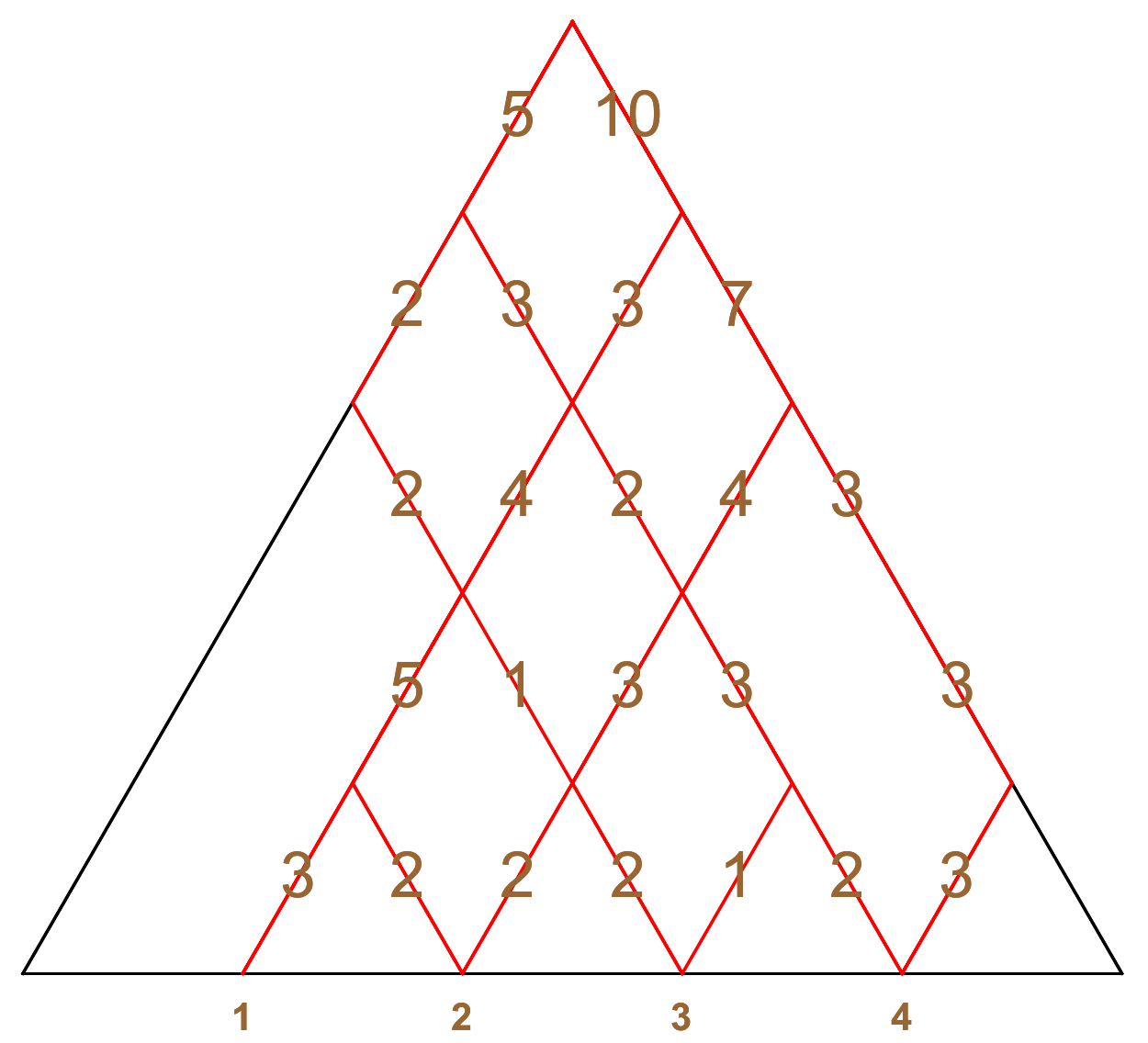}
       \vspace{0.5 cm}
       \includegraphics[scale=0.5]{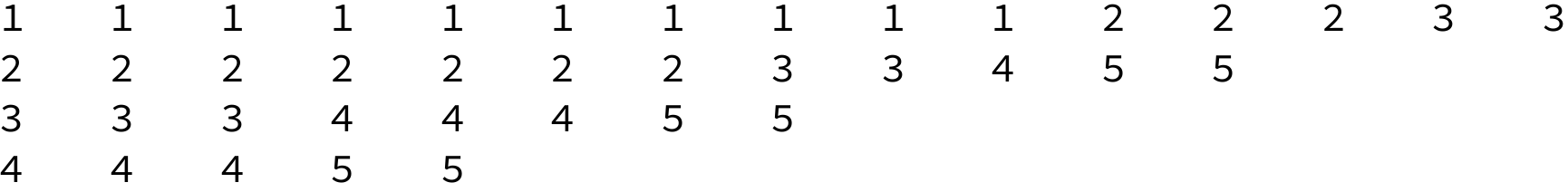}
         \hspace{0.8 cm}
         \includegraphics[scale=0.5]{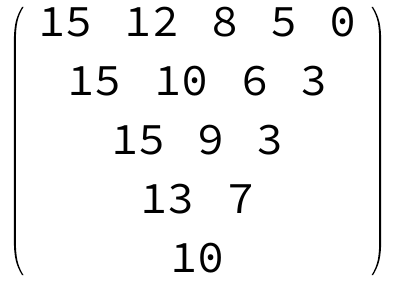}
    \caption{\label{lianasExample} An example of a set of lianas (Left), whose superposition (Right) builds a liana forest to be identified (zoom in!) with the reduced O-blade of Fig.~\ref{fullobladeforkostka}. It is associated with a semi-standard Young tableau (Bottom-L) and with a Gelfand-Tsetlin pattern (Bottom-R). } 
    \end{figure}

Lianas and liana forests, like the O-blades, were discovered by \cite{Ocn},  
Several aspects of this construction have been recently summarized in a video lecture \cite{AO:lianasHarvard} and some details can be found in the appendix of our paper \cite{CZ:Kostka} (the zig-zag lines are sometimes called ``wires'' in \cite{AO:lianasHarvard}, the terminology
``lianas'' and ``liana forest'' was introduced in \cite{CZ:Kostka}).
One example is displayed on Fig.~\ref{lianasExample}.
We cannot give an account of this theory here and we suggest the interest reader to look at the quoted references.

 \section{Semi-classical limit of multiplicities}
 \subsection{About Horn and Schur volume functions (generalities)}
 \label{sec:volumefunctions}

When the highest weights of the three irreps $\lambda, \mu, \nu$ are scaled by a non-negative integer $s$, the obtained multiplicities, in the case of $\SU(n)$,  are given by a polynomial\footnote{For simple Lie groups other than $\SU(n)$, it can be a quasi-polynomial, \cite{DW}, \cite{Rass}.} in $s$ (stretching polynomial, or LR polynomial:
$C_{s\lambda, s\mu}^{s\nu}= {\mathcal J}(\lambda, \mu, \nu) \, s^d + \ldots$)  in which the leading coefficient, in the generic case, is given by the so-called {\it Horn volume function} denoted ${\mathcal J}(\lambda, \mu, \nu)$; this function can be considered, in the terminology of geometric quantization, as a semi-classical limit of LR-multiplicities. 

The so-called {\it Schur volume function}, denoted ${\mathcal I}(\lambda, \delta)$, plays the same role as ${\mathcal J}(\lambda, \mu, \nu)$ in the second type of problem involving multiplicities, \ie when one scales both the highest weight of an irrep and some other weight of its weight system, \ie when one consider multiplicities $K_{s\lambda, s\delta}$.

Because of Eq.~\ref{LRversusKostka}, the two problems, Horn versus Schur, like LR-coefficients versus Kostka numbers, are not independent either.

Both volume functions can actually be defined for arbitrary general arguments, \ie one may consider real, but non necessarily integral weights:\\
Up  to a simple pre-factor the volume function ${\mathcal I}$ is the PDF (probability density function) that appears in the Schur problem :  it describes the distribution of the diagonal elements of a Hermitian matrix drawn randomly and uniformly on a $\SU(n)$ orbit  (the group acting by conjugation on the space of Hermitian matrices) specified by a set of ordered eigenvalues. 
The same volume function appears as the inverse Fourier transform of an orbital integral identified with the density of a Duistermaat-Heckman's measure.\\
In the same way, the volume function ${\mathcal J}$, up to appropriate prefactors, is the PDF that appears in the Horn problem : it describes the distribution of the eigenvalues of a sum of two Hermitian matrices, both drawn randomly and uniformly on two $\SU(n)$ orbits\footnote{This situation can obviously be generalized (choice of the Lie group, choice of the space on which it acts). We refer to \cite{CZ2}, \cite{CMSZ}, for a study of several various cases.}. 
This PDF appears as inverse Fourier transformation of the product of Fourier transforms of two measures, it involves three copies of the same orbital integral  (intuitively, it can be thought of as an orbital analog of $\delta_{a+b}(c)$,  where $\delta$ is the Dirac measure). 

We have introduced $\SU(n)$ hive polytopes for any triple $(\lambda,\mu,\nu)$ of compatible highest weights in the dominant Weyl chamber that are such that $C_{\lambda,\mu}^\nu \neq 0$ ({\sl compatible} means that $\lambda+\mu-\nu$ should belong to the root lattice, this is a necessary condition for $C_{\lambda\mu}^\nu$ not to vanish), in particular they are integral points. When the dimension of  such a hive polytope is equal to $d=(n-1)(n-2)/2$, \ie for a generic triple, the volume function ${\mathcal J}$ evaluated at the same arguments indeed gives the (Riemannian) volume of this polytope ---see our example in the next section. This result justifies the terminology ``volume function''. 
Notice that the system of equations and  inequalities defining hive polytopes depend linearly on the three arguments, so that one can extend their definition and consider hive polytopes for any triple of (real) points. 
This provides another way to define the ${\mathcal J}$ volume function for real arguments\footnote{When using real arguments, it is preferable to use Young variables (which become partitions, in the case of integral weights) rather than Dynkin components.}.
There are similar considerations and results for the Schur volume function ${\mathcal I}$.

It can be shown that volume functions are real-analytic away from a particular collection of hyperplanes.
Non-analyticities, of ${\mathcal J}$ for instance, arise from changes in the geometry of the hive polytope as the third argument varies (the first two being held constant).
${\mathcal J}$ is a piecewise polynomial function on its third argument and the non-analyticities take the form of a change of polynomial determination. Again there are similar considerations for ${\mathcal I}$.
It is interesting to investigate the differentiability class of the volume functions, as well as the nature of singularities, but we shall say nothing about this problem here, 
we only notice that this is one of the subjects studied in \cite{CZ2},  \cite{CZ:Kostka},  \cite{CMSZ}, and we refer the interested reader to these articles.
The other aspects of ${\mathcal I}$ and ${\mathcal J}$ briefly mentioned in the previous paragraphs, in particular their relations with the Horn  and the Schur problems, will not be discussed in the present paper either, we refer 
 the reader to the same articles, as well as to \cite{CZ1, CZ3, McS:Horn}, and to the forthcoming thesis \cite{McS:thesis}.
 
  \subsection{Volume functions and $R$-polynomials}
  \label{sec:Hornvolumefunction}

In this paper we shall only discuss and illustrate two aspects of the general theory. 
One is the fact that these volume functions can be obtained (or defined, since we shall limit ourselves to the case of integral arguments) by looking at the dominant term of the polynomials obtained from the multiplicities (LR or Kostka) by scaling the weights: this will be the subject of the next section.
The other is a property relating the Horn volume function ${\mathcal J}$ to characters of $\SU(n)$:

Let us first remind the reader the expression of the LR multiplicity $C_{\lambda\mu}^\nu$ as the integral of a product of characters $\chi_\lambda \chi_\mu \chi_\nu^*$ over the Cartan torus $\T_n \,= \U(1)^{n-1}$  of $\SU(n)$:
$$
C_{\lambda\mu}^\nu =\int_{ \T_n} dT\, \chi_\lambda(T) \chi_\mu(T) \chi_\nu^*(T)
$$
where $T=\text{diag}(e^{\ii t_j})$ and $dT$ is the normalized Haar measure on $ \T_n$. 
There is\footnote{This was shown for $\SU(n)$ in \cite{CZ1} and generalized to all semi-simple Lie algebras in \cite{CMSZ}.} an analogous expression for the semi-classical limit ${\mathcal J}(\lambda, \mu, \nu)$, it reads as follows:
if $(\lambda,\mu,\nu)$ is a compatible triple of highest weights and if we denote by primes $\lambda',\mu',\nu'$  the shift of those weights by the Weyl vector $\rho$, i.e. $\lambda'=\lambda+\rho$, etc., one has
\begin{subequations}
\label{Rpolynomials}
\begin{align}
{\mathcal J}(\lambda',\mu', \nu') &=  \int_\T dT \;  R(T) \; \chi_\lambda(T) \chi_\mu(T)  (\chi_\nu(T)^* \\
{\mathcal J}(\lambda,\mu, \nu) &= \int_\T dT  \;  \hat R (T) \;   \chi_{\lambda-\rho}(T) \chi_{\mu-\rho}(T)  (\chi_{\nu-\rho}(T) )^*
\end{align}
\end{subequations}
where the two functions $R(T)$ and $\hat R(T)$ are real multivariate invariant\footnote{More generally they are invariant under the underlying Weyl group, which, for $\SU(n)$, is the symmetric group.} polynomials of the $e^{\ii t_j}$ (using appropriate variables they are trigonometric polynomials in $n-1$ real variables) depending only on the choice of the Lie group, here $\SU(n)$.
There exists a finite, $n$-dependent set $\CK$ of dominant weights such that $R_n(T)$ can be written as  
 a  linear combination $R_n(T)=\sum_{\kappa \in \CK}  r_\kappa \chi_\kappa(T)$ of real characters; the coefficients $r_\kappa$ are non-negative rational numbers and are such that, when evaluated at the identity matrix, $R_n(\mathds{1})=1$.
There is a similar theorem for $\hat R_n(T)$, with another set  $\widehat{\CK}$ of dominant weights, and with new coefficients $\hat  r_\kappa$.
These two theorems were shown for $\SU(n)$ in \cite{CZ1}, and generalized to arbitrary semi-simple Lie groups in \cite{ER}.
The  functions $R$ and $\hat R$ can be calculated as iterated series and their explicit expressions for $SU(n)$, as well as their decompositions on characters, are given in \cite{CZ1}, for $n=3,4,5,6$.
In \cite{ER} they are called Mittag-Leffler type sums associated with root systems and the authors provide a general description of the sets $\CK$ and $\widehat{\CK}$, \ie of the weights that enter the decomposition of the two functions on characters (we shall come back to this in section \ref{Rgeneral} and in the appendix).  In every particular case, \ie for each choice of a semi-simple Lie group, the a priori knowledge of these two sets is a very useful piece of information, but the explicit determination of the coefficients $r_\kappa$ and $\hat  r_\kappa$ remains nevertheless a difficult task.
Since these coefficients are known for $\SU(n)$, $n\leq6$, from the work of \cite{CZ1}, we shall here consider the next case: in section \ref{sec:R7}  we give the expression of $R_7 = \hat R_7$ (the two functions are indeed equal for $\SU(n)$ when $n$ is odd) and explain the method -- this particular result (as well as a few comments in the appendix) is actually the only original contribution of the present paper, the rest being, partly a review, partly an attempt to provide a pedagogical introduction to the subject.
 As a warm-up we first re-consider the case $n=5$,  which is much simpler. 
  
 \subsection{LR-polynomial and the Horn volume function: an $\SU(5)$ example}
 \label{sec:stretchedLR}
  
 Let us consider the multiplicities obtained by scaling the three highest weights of $\SU(5)$ considered in the previous section, namely $\lambda=\{3,4,3,5\}$,  $\mu=\{4,3,5,4\}$, $\nu=\{2,2,4,2\}$. 
 We know that $C_{\lambda, \mu}^{\nu}=371$. We find  $C_{s\lambda, s\mu}^{s\nu} =  371, 7983, 60849, 277394, 930849, 2548764, 6037641$ for $s=1,2,\ldots, 7$.
 Going up to $s=7$ is necessary to determine the LR-polynomial because we a priori know that it is of degree $d=6$ (the triple is generic and the degree is the number of inner vertices of $\SU(5)$ O-blades, aka the dimension of the hive polytope).
 We also know that it is a polynomial (not a quasi-polynomial) because the Lie group is of type $\SU(n)$, and because the chosen triple is compatible (this is implied by the fact that  $C_{\lambda, \mu}^{\nu}\neq 0$).
 The obtained interpolating polynomial is 
 \[ \frac{314 s^6}{9}+\frac{11593 s^5}{120}+\frac{8401 s^4}{72}+\frac{1921 s^3}{24}+\frac{2407 s^2}{72}+\frac{167 s}{20}+1\]
 One can then use this LR-polynomial to obtain for instance the multiplicity of the branching rule $\lambda=\{300,400,300,500\}$, $\mu=\{400,300,500,400\}$, $\nu=\{200,200,400,200\}$, \ie choosing a scaling factor $s=100$; one finds $C_{\lambda, \mu}^{\nu}=35866720654586 = 2 \times 41\times 143387 \times 3050479$, a result that would be totally out of reach by a direct attempt.
 
 The value of the Horn volume function is given by the dominant term, therefore one finds ${\mathcal J}(\{3,4,3,5\}, \{4,3,5,4\}, \{2,2,4,2\}) = \frac{314}{9}$.
 Of course we could have obtained this value by plugging the chosen highest weights in the explicit expression of the function ${\mathcal J}$ since this function is known for $\SU(5)$, see \cite{CZ1}, but we do not want to use this information here.
 Another way to determine the above value could have been to calculate directly the euclidean volume of the hive polytope since it is given by the dominant coefficient of the Ehrhart polynomial of the polytope, which coincides in the present case with the LR-polynomial.
 
Calculating multiplicities for scaled weights (what we did above to determine the LR-polynomial) is obviously a painful operation that can be done, at least for the previous example, by using available dedicated computer algebra packages, like LiE \cite{LiE}, or by writing algorithms based on the theory of O-blades and honeycombs  discussed in the last section; in the present paper we used Mathematica\footnote{All the examples discussed in this paper (graphics, multiplicities, $\ldots$) and many algebraic manipulations have also been worked out with the help of Mathematica. Packages for O-blades and metric honeycombs are available on the web site \cite{RC:packages}.} \cite{Mathematica} to write such algorithms. 
Calculating ${\mathcal J}(\lambda, \mu, \nu)$ in this way for a compatible triple already breaks down, unfortunately, for the $n=7$ examples considered in the next section, because one should then calculate multiplicities for scaled arguments using scaling factors going up to $s=15+1$, and this exhausts the memory resources of most computers.

  \subsection{The $R$-polynomials of $\SU(n)$}
  \label{Rgeneral}
  Here we explain how to determine the $R$-polynomials that appear in eq.~\ref{Rpolynomials}. We know of two (plus one) ways to compute them. The first is a brute force calculation involving iterated series, this was the method used in \cite{CZ1}. In the same reference it was observed that $R$ and $\hat R$ can be expressed as a finite sum of characters over a set $\CK$, resp. $\hat \CK$ of highest weights. This observation was later proved in full generality, \ie for all semi-simple Lie algebras,  in \cite{ER}, where the set $\CK$ was precisely identified as the set of weights belonging to the \underline{interior} of the convex hull of the Weyl group orbit of the Weyl vector $\rho$ (more about it in the appendix). This gives rise to a second method since it allows one to write an a priori decomposition of $R$ with unknown coefficients $r_\kappa$; there is an analogous decomposition for $\hat R$. The coefficients $r_\kappa$ (and $\hat r_\kappa$) still need to be determined; the idea is then to use the following equations (from \cite{CZ1}, \cite{CMSZ}), relating these unknown coefficients to the volume functions themselves :
 \be
 \label{kissinger} \c_\kappa= \CJ(\rho,\rho,\kappa+\rho) {,\ \kappa\in \CK\qquad\mathrm{and} \qquad \hat\c_\kappa=\CJ(\rho,\rho,\kappa+\rho) ,\ \kappa\in \hat \CK }  \,.
 \ee
 Even in the cases where the volume functions are not explicitly known, one can in principle determine the LR-polynomial of the weights and read their values for those particular arguments. The drawback of this approach is that it becomes unpractical if the chosen example is such that the determination of the LR-polynomials exhausts the memory resources (this will be the case when $n=7$). There is a third method, a kind of mixture of the first two, that can be applied for $n=7$ ---more about it later.
 Let us give some more details about the first two methods and discuss the example $n=5$. 
 
$\bullet$ First method. For general $\SU(n)$, equations 25-27 and 29-30 of \cite{CZ1} express $R$ respectively $\hat R$ as iterated series (Mittag-Leffler type sums). 
 We give them here again  for the convenience of the reader; the expression of $\hat R$ is similar (there is no sign factor in front of the $\prod$ symbol); in the examples $n=5,7$ that we shall discuss below, both are anyway equal (because $n$ is odd).\\
 Assuming that $T=\text{diag}(e^{i t_j})$ is a unitary unimodular matrix, hence $\sum t_j = 0$, and setting  $u_j=t_j - t_{j+1}$, one finds:
  \begin{equation}
 \begin{aligned}
 R_n(T) &= D_n \varpi_n \quad \text{where} \\
 D_n &= \sum_{p_1, \cdots, p_{n-1} =-\infty}^\infty (-1)^{\sum_j j p_j (n-1)} \prod_{1\le i < i'\le n}   \frac{1}{u_i+u_{i+1}+\cdots +u_{i'-1}+(p_i+\cdots +p_{i'-1})(2\pi)} \\
 \varpi_n&= \prod_{1\le i < i'\le n} 2 \sin(\oh(u_i+u_{i+1}+\cdots +u_{i'-1}))
 \end{aligned}
 \label{eq:iteratedseries}
\end{equation}
Evaluation of the above iterated sum leads to a complicated real trigonometric polynomial in the $n-1$ variables $u_j$ or in the $n$ variables $t_j$.
The last step is to recognize that the whole expression can be recast in a simpler form, by using the Weyl characters $\chi$ of $\su(n)$. 
 This is how $R_n$, $n=2\ldots 6$, was first calculated\footnote{The case $n=2$ requires a special treatment because the sum (\ref{eq:iteratedseries}) diverges for this value.} (in \cite{CZ1}).
In the case $n=5$, for example, one finds:
 \begin{equation}
R_5 = {\hat R_5}  = \frac{1}{360} (45 + 10 \chi_{1,0,0,1} + \chi_{0,1,1,0})
\label{R5result}
 \end{equation}

$\bullet$ Second method.  We consider the interior of the convex hull of the group orbit of the Weyl vector \cite{ER}.
For $n=5$ it contains only the weights $\{0,0,0,0\}$, $\{1,0,0,1\}$ and $\{0,1,1,0\}$. One obtains therefore immediately a decomposition of $R_5$ of the type given by eq.~\ref{R5result}. The coefficients remain now to be found.
Using O-blades or the program LiE, together with formulae \ref{kissinger}, one finds, under scaling by an integer $s=0\ldots 7$, the following first seven multiplicities (actually eight, the last being used as a test):\\  
{\small
$C_{s\{1,1,1,1\}, s\{1,1,1,1\}}^{s\{1,1,1,1\}} = 1,16,126,616,2200, 6336,15631,34336 $ leading to the polynomial\\ $1+(13 s)/4+(37 s^2)/8+4 s^3+(9 s^4)/4+(3 s^5)/4+s^6/8$\\
$C_{s\{1,1,1,1\}, s\{1,1,1,1\}}^{s\{2,1,1,2\}} = 1,12,74,304,959,2520,5796,12048$ leading to the polynomial\\ $1+(89 s)/30+(34 s^2)/9+(11 s^3)/4+(43 s^4)/36+(17 s^5)/60+s^6/36$\\
$C_{s\{1,1,1,1\}, s\{1,1,1,1\}}^{s\{1,2,2,1\}} = 1,8,35,112,294,672,1386,2640 $ leading to the polynomial\\ $1+(157 s)/60+(949 s^2)/360+(4 s^3)/3+(13 s^4)/36+s^5/20+s^6/360$\\}
One recognize the coefficients $r_\kappa=1/8, 1/36, 1/360$ of $R_5$ from the leading terms of the above three polynomials.
 
$\bullet$ Third method. See \ref{sec:R7}.
 
   \subsubsection*{Alternative expressions for $R$-polynomials}
   
   Given a partition $p=[p_1, p_2, \ldots, p_{n-1}]$, one can obtain the Lie group Weyl character of  $\SU(n)$ from the Schur polynomial $s_p$ in $n$ variables $X_1,X_2,\ldots X_{n}$ by setting $X_1=y_1, X_2=y_2/y_1, \ldots, X_j=y_j/y_{j-1}$ and imposing $y_n=X_1X_2\ldots X_n=1$; it is a Laurent polynomial in the $y_1,\ldots, y_{n-1}$.  
   A Lie algebra Weyl character of $\su(n)$ is obtained from the same Schur polynomial $s_p$ by setting $X_j = exp(\ii t_j)$, while imposing $t_j = \theta_j - \tfrac{1}{n} \sum_i \theta_i$ (therefore $\sum t_j = 0$), and finally $\theta_j = \theta_1 - \sum_{i=1}^{j-1} u_j$, equivalently $u_j = \theta_j - \theta_{j+1} = t_j-t_{j+1}$ for $j=1,\ldots, n-1$; it is a trigonometric polynomial in the variables $u_1, \ldots, u_{n-1}$, the same as those used in eq.~\ref{eq:iteratedseries}.
   
Since the $R$ polynomials can be expanded on the Weyl characters $\chi_\lambda$ of $\su(n)$ and since the latter are simply related (as recalled above) to Schur polynomials ($s_{\ell(\lambda)}$) where $\ell(\lambda)$ is the Young partition with components\footnote{We denote partitions with braces, for instance $\ell{(\{1,0,0,1\})}=[2,1,1,1,0].$} $\ell_i(\lambda)=\sum_{j=i}^{n} \lambda_j$ and $\lambda_j$ are Dynkin coordinates, with $\lambda_n=0$, it is natural to see if another linear basis of the space of symmetric polynomials would lead to a further simplification of $R$, or $\hat R$.
 Expressing Schur polynomials ($s_p$) in terms of other families, for instance elementary symmetric polynomials ($ {e} _ j$), monomial symmetric polynomials ($m_p$), or power-sum symmetric polynomials ($ {p} _ j$), we find, for the $R_5$ example of eq.~\ref{R5result}, the following alternative expressions: $ \frac{1}{36} s_{[2,1,1,1,0]}+\frac{1}{360} s_{[2,2,1,0,0]}+\frac{1}{8}$, 
  $\frac {1} {360}\left ( {e} _ 2 {e} _ 3 +  9 {e} _ 1 {e} _ 4 - 10 {e} _ 5 + 45 \right)$, $\frac{1}{36} (4 {m}_{[1,1,1,1,1]}+{m}_{[2,1,1,1,0]})+\frac{1}{360} (5 {m}_{[1,1,1,1,1]}+2 {m}_{[2,1,1,1,0]}+{m}_{[2,2,1,0,0]})+\frac{1}{8}$,  
$\frac{1}{2880} \, ({3 {p}_1^5-14 {p}_2 {p}_1^3+12 {p}_3 {p}_1^2+\left({p}_2^2+2 {p}_4\right) {p}_1+4 \left(3 {p}_2 {p}_3-4 {p}_5+90\right)})$.
Clearly, the expansion on Schur polynomials looks nicer, and it appears to be more natural in this framework.

 \subsubsection*{Using the $R$-polynomial: from LR-coefficients to the Horn volume function (and back)}
   If we plug the decomposition of $R_n$ on Lie group characters, for example equation \ref{R5result}, into the expression~\ref{Rpolynomials} giving the volume functions in terms of $R_n$ (or $\hat R_n$) and characters, 
   one obtains ${\mathcal J}(\lambda, \mu, \nu)$, or the same function with $\rho$-shifted arguments, as a local average of a finite number of LR-coefficients. Assuming that  $(\lambda,\mu;\nu)$ is a compatible triple, and still denoting the $\rho$-shifted weights by prime indices,  one finds (see \cite{CZ1} and \cite{CMSZ}):

 \be {\mathcal J}(\lambda',\mu',\nu')  = \sum_{\kappa\in \CK, {\tau}} \c_\kappa  C_{\lambda\,\mu}^{{\tau}}  C_{{\tau}\,\kappa}^{{\nu}}, \qquad
   {\mathcal J}(\lambda,\mu,\nu) =  \sum_{\kappa\in \hat \CK, {\tau}} \hat\c_\kappa  C_{(\lambda-\rho)\,(\mu-\rho)}^{{\tau}}  C_{{\tau}\,\kappa}^{{\nu-\rho}}
\ee

 Several examples are worked out in Ref.~\cite{CZ1}. 
Inverting the above formulae (and their analog for the Schur volume function) is a very interesting question that has been addressed in \cite{McS:Horn}.

 \subsection{The expression of $R_n$ when $n=7$ (a new result)}
 \label{sec:R7}
 Using method (1) of section  \ref{Rgeneral} only partially works because, although the summation of the iterated series can explicitly be done,  it leads to an unprintable huge expression involving trigonometric lines, out of which it is very difficult to read a decomposition on characters.
 Using method (2) of section  \ref{Rgeneral} also partially works: the characters that appear in the expression of $R_7$ are readily determined (see below), but calculating the associated coefficients by following the technique used in the example $R_5$ cannot be done because the calculation of multiplicities for scaled weights exhausts the memory resources of most available computers (here the degree of LR-polynomials, in the generic case, is $d=15$, and it turns out that one has to calculate LR coefficients involving weights as large as $\rho + s \kappa$ with $s = 1,2,\ldots 15,16$, with weights $\kappa$ as in eq.~\ref{R7result}).
The idea (method 3) is a mix of these two methods:
We first determine the dominant weights whose characters should occur in $R_7$ by finding the dominant weights that belong to the interior of the convex hull of the Weyl orbit of the Weyl vector, we then compute the Lie algebra characters associated with those weights.
We obtain in this way a decomposition of $R_7$ as a trigonometric polynomial in six variables involving $21$ unknown coefficients; some of the weights (6 of them) are of complex type and appear with their complex conjugate, the coefficients $r_\kappa$ associated with such pairs have to be equal since the final expression is real, this brings the number of unknown coefficients to $15$. From the other hand we perform a brute force calculation of $R_7$ by summing the series eq.~\ref{eq:iteratedseries}, the obtained expression is quite large (not more than 40 pages) but still manoeuvrable.  We then identify the two obtained expressions ---making sure that they use the same set of variables!--- by choosing particular values for the six angular variables; care has also to be taken at this point because several simple choices lead to expressions containing individual terms with vanishing denominators, such terms should of course disappear after proper simplification. This identification leads to a linear system in the variables $r_k$ which is of maximal rank ($15$) for good choices of the chosen set of particular angular values. The last step is to solve this system. Here is the result:

\small{
\begin{equation}
\begin{split}
R_7 = {\hat R_7}  =
\frac{1}{3\times13!} &
(
   87766794 +
    2 \, \chi_{0, 1, 1, 1, 1, 0} + 
    29 (\, \chi_{0, 1, 2, 0, 0, 1} + \, \chi_{1, 0, 0, 2, 1, 0}) + 
    38 \, \chi_{0, 2, 0, 0, 2, 0} +\\ &
    647 (\, \chi_{0, 2, 0, 1, 0, 1}+ \, \chi_{1, 0, 1, 0, 2, 0}) +
    3575 (\, \chi_{0, 0, 0, 1, 2, 0} + \, \chi_{0, 2, 1, 0, 0, 0}) +
    13188 \, \chi_{1, 0, 1, 1, 0, 1} +\\ &
    75599 (\, \chi_{0, 0, 0, 2, 0, 1} + \, \chi_{1, 0, 2, 0, 0, 0}) +
    88248 \, \chi_{1, 1, 0, 0, 1, 1} +
    313706 \, \chi_{2, 0, 0, 0, 0, 2} +\\ &
    554727 (\, \chi_{0, 0, 1, 0, 1, 1} + \, \chi_{1, 1, 0, 1, 0, 0}) +
    2157704 (\, \chi_{0, 1, 0, 0, 0, 2} + \, \chi_{2, 0, 0, 0, 1, 0}) +\\ &
    3601542 \, \chi_{0, 0, 1, 1, 0, 0} +
    15350862 \, \chi_{0, 1, 0, 0, 1, 0} +
    46669412 \, \chi_{1, 0, 0, 0, 0, 1})
    \label{R7result}
  \end{split}
\end{equation}
    }
       
 Using the dimensions\footnote{Using the same order of appearance as in eq.~\ref{R7result} these dimensions are: \\ ${1, 105840, 2\times 30870, 27000,  2\times 26460, 2\times 3528, 24500, 2\times 2646, 10240, 735, 2\times 2940, 2 \times 540, 784, 392, 48}$.} 
 $\dim V_\kappa$ of the representations spaces of highest weights $\kappa$ that label the characters of the decomposition of $R_7$, one can check the self-consistency of the obtained result for their coefficients $r_\kappa$, in eq.~\ref{R7result},  by verifying the identity  $\sum_{\kappa} r_\kappa \dim V_\kappa=1$,  in agreement with the constraint $R_n(\mathds{1}) = 1$.
 
 \subsection{Stretched multiplicities and the Schur volume function}

 In Sec.~\ref{sec:Hornvolumefunction} and \ref{sec:stretchedLR} we choose to discuss the Horn volume functions in relation with stretched Littlewood-Richardson multiplicities. 
 For lack of time and space we  skip the corresponding  discussion about stretched Kostka numbers (see \cite{McAllister:Kostka}) and about the Schur volume function.
 This last topic could be handled in a  way similar to the previous discussion and we refer the reader to reference \cite{CZ:Kostka}.

 \section{Appendix}
 
For a general simple or semi-simple Lie algebra, let us call $(\bm{\alpha}_j)$ a basis of simple roots and $(\bm{\omega}_j^\vee)$ the dual basis of fundamental co-weights.
According to \cite{ER}, in order for the character $\chi_\kappa$ to occur in $R_n =\sum_{\kappa \in \CK}  r_\kappa \chi_\kappa$, the weight $\kappa$ should belong to the interior of the convex hull of the Weyl group orbit of $\rho$.\\
Equivalently, if the Weyl vector $\rho$ belongs to the root lattice $Q$, then $R = \hat R$; define $\xi$ as the sum of simple roots $\bm{\alpha_j}$.
If, on the contrary, $\rho \notin Q$, then $R \neq \hat R$;  define $\xi$ (resp.~$\hat \xi$) from $\sum \bm{\alpha}_j$ by replacing each $\bm{\alpha_j}$ by $\tfrac{1}{2} \bm{\alpha_j}$ whenever  $\langle \rho, \bm{\omega}_j^\vee\rangle$  \underline{is not} an integer (resp.~whenever  $\langle \rho, \bm{\omega}_j^\vee\rangle$  \underline{is} an integer), 
then $\CK$ (resp.~$\hat \CK$) is  the set of dominant weights belonging to the weight system of the irrep of highest weight $\mathring{\rho} \doteq \rho-\xi$ (resp.~$\mathring{\hat \rho} \doteq  \rho-\hat\xi$).

$R = \hat R$ in the following cases:  $A_r$ with $r$ even, $D_r$ with $r=4c$ or $r=4c+1$ where $c$ is a positive integer, $E_6$, $E_8$,  $G_2$, $F_4$, and $C_r$ with $r=4c+3$ or $r=4c+4$ where  $c$ is a non-negative integer.
 Notice that it is so, in particular, for $\SU(n)$ with $n$ odd.

$R \neq \hat R$ in the following cases:  $A_r$ with $r$ odd,  $D_r$ with $r=4c+2$ or $r=4c+3$ where $c$ is a positive integer, $E_7$,  $B_r$, and $C_r$ with $r=4c+1$ or $r=4c+2$ where  $c$ is a non-negative integer.
 Notice that it is so, in particular, for $\SU(n)$ with $n$ even.
 
For illustration, we list the sets $\CK$ and $\hat \CK$ for several Lie algebras\footnote{In the non simply-laced cases we introduce a semi-colon to single out short roots from long roots in Dynkin components, short roots are to the right in all cases, with the exception of $C_r$.} of small rank (for exceptional Lie groups of larger ranks we only give the highest weights  $\mathring{\rho}$ and  $\mathring{\hat \rho}$ and their Dynkin components):

\smallskip
{\scriptsize

$A_1$. $\CK= \{\{0\}\}$. $\qquad$ $\hat \CK= \{\{1\}\}$.

$A_2$.  $\CK = \hat \CK =  \{\{0, 0\}\}$.

$A_3$. $\CK = \{\{1, 0, 1\}, \{0, 0, 0\}\}$. $\qquad$ $\hat \CK= \{\{0,1,0\}\}$.

$A_4$. $\CK = \hat \CK = \{ \{0, 1, 1, 0\}, \{1, 0, 0, 1\}, \{0, 0, 0, 0\}\}$.

$A_5$. $\CK = \{\{1, 0, 2, 0, 1\}, \{1, 1, 0, 1, 1\}, \{0, 0, 1, 1, 1\}, \{1, 1, 1, 0, 
  0\}, \{2, 0, 0, 0, 2\}, \{0, 1, 0, 0, 2\}, \{2, 0, 0, 1, 0\},
   \{0, 0, 2, 0, 0\}, \{0, 1, 0, 1, 0\},\\ \{1, 0, 0, 0, 1\}, \{0, 0, 0, 0, 0\}\}$. $\qquad$
$\hat \CK=  \{\{0, 0, 1, 0, 0\}, \{1, 1, 0, 0, 0\}, \{0, 0, 0, 1, 1\}, \{1, 0, 1, 0, 1\}, \{0, 2, 0, 0, 1\}, \{1, 0, 0, 2, 0\}, \{0, 1, 1, 1, 0\}\}$.

$A_6$. $\CK = \hat \CK = \{\{0, 1, 1, 1, 1, 0\}, \{1, 0, 0, 2, 1, 0\}, \{0, 2, 0, 0, 2, 0\}, \{0, 1, 2,
   0, 0, 1\}, \{1, 0, 1, 0, 2, 0\}, \{0, 2, 0, 1, 0, 1\}, \{1, 0, 1, 1, 0,  1\},\\  \{1, 1, 0, 0, 1, 1\}, 
   \{0, 0, 0, 1, 2, 0\}, \{0, 2, 1, 0, 0, 0\}, \{1, 0, 2, 0, 0, 0\}, \{0, 0, 0, 2, 0, 1\}, \{1, 1, 0, 1, 0, 0\}, \{0, 0, 1, 0, 1, 1\},\\ \{0, 0, 1, 1, 0, 0\}, \{2, 0, 0, 0, 0, 2\}, \{0, 1, 0, 0, 0, 2\}, 
  \{2, 0, 0, 0, 1, 0\}, \{0, 1, 0, 0, 1, 0\}, \{1, 0, 0, 0, 0, 1\}, \{0, 0, 0, 0, 0, 0\}\}$.
  
 $D_4$. $\CK = \hat \CK = \{ \{0, 2, 0, 0\}, \{1, 0, 1, 1\}, \{0, 0, 0, 2\}, \{0, 0, 2, 0\}, \{2, 0, 0, 
  0\}, \{0, 1, 0, 0\}, \{0, 0, 0, 0\} \}$. 
  
  $D_5$. $\CK = \hat \CK = \{ \{0, 1, 2, 0, 0\}, \{0, 2, 0, 1, 1\}, \{1, 0, 1, 1, 1\}, \{1, 1, 0, 0, 2\}, \
\{1, 1, 0, 2, 0\}, \{0, 3, 0, 0, 0\}, \{1, 1, 1, 0, 0\}, \{0, 0, 0, 2, 2\}, \\
\{0, 0, 1, 0, 2\}, \{0, 0, 1, 2, 0\}, \{0, 0, 2, 0, 0\}, \{2, 0, 0, 1, 1\}, \
\{0, 1, 0, 1, 1\}, \{2, 1, 0, 0, 0\}, \{0, 2, 0, 0, 0\}, \{1, 0, 0, 0, 2\}, 
\{1, 0, 0, 2, 0\},\\ \{1, 0, 1, 0, 0\}, \{0, 0, 0, 1, 1\}, \{2, 0, 0, 0, 0\}, \{0, 1, 0, 0, 0\}, \{0, 0, 0, 0, 0\} \}$.

$G_2$. $\CK = \hat \CK = \{ \{0; 2\}, \{1; 0\}, \{0; 1\}, \{0; 0\} \}$.

$B_2$. $\CK = \{\{1; 0\}, \{0; 0\}\}$. $\qquad$ $\hat \CK = \{\{0; 1\}\}$.

$B_3$. $\CK = \{  \{0, 0; 0\}, \{1, 0; 0\}, \{0, 1; 0\}, \{2, 0; 0\}, \{0, 0; 2\}, \{1, 1; 0\}, \{1, 0; 2\}  \}$. $\qquad$ $\hat \CK=  \{\{0, 0;1\}, \{1, 0; 1\},\{0, 1; 1\}\}$.

$C_3$.  $\CK = \{\{0, 2; 0\}, \{1, 0; 1\}, \{2, 0; 0\}, \{0, 1; 0\}, \{0, 0; 0\}\}$.

$F_4$.  $\mathring{\rho}=-\sum \bm{\alpha_j}+\rho = \{0, 1; 2, 0\}$.

$E_6$.   \includegraphics[width=6pc,height=1pc]{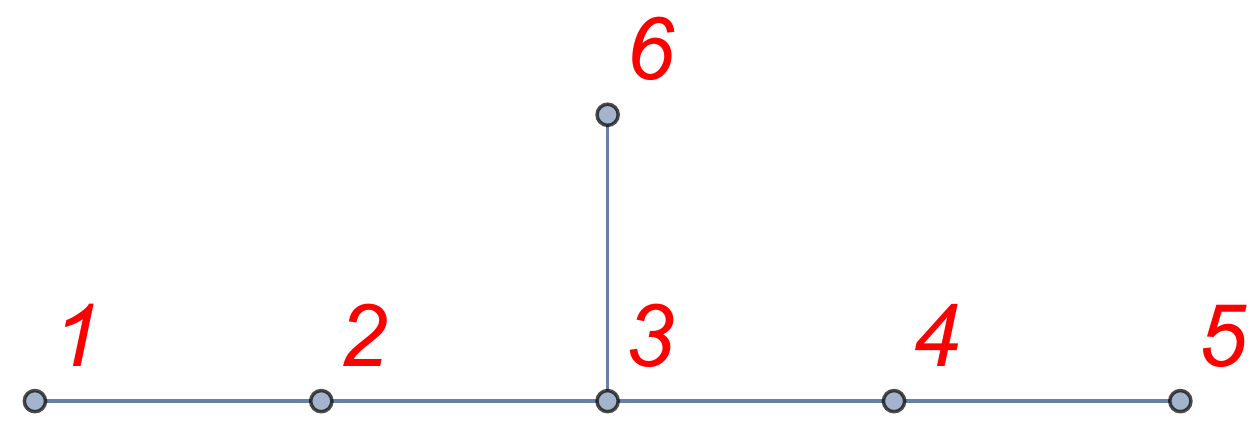}  $\mathring{\rho}=-\sum \bm{\alpha_j}+\rho$=$\{0, 1, 2, 1, 0, 0\}$.
 
$E_7$.   \includegraphics[width=6pc,height=1pc]{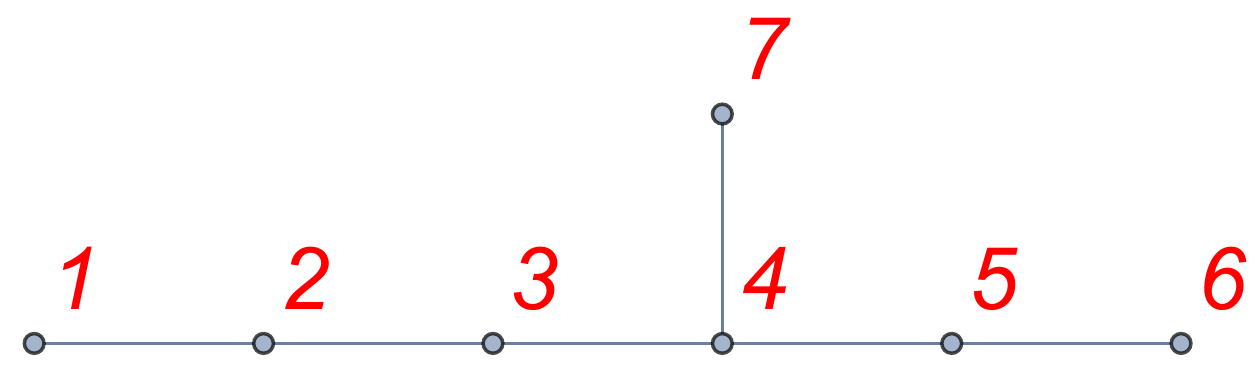} $\mathring{\rho}=-\frac{\alpha _1}{2}-\alpha _2-\frac{\alpha _3}{2}-\alpha _4-\alpha _5-\alpha _6-\frac{\alpha_7}{2}+\rho = \{1, 0, 2, 1, 1, 0, 1\}$.  \;  $\mathring{\hat \rho} = -\sum \bm{\alpha_j}+\rho=\{0, 1, 1, 2, 1, 0, 0\}$. 

$E_8$.    \includegraphics[width=6pc,height=1pc]{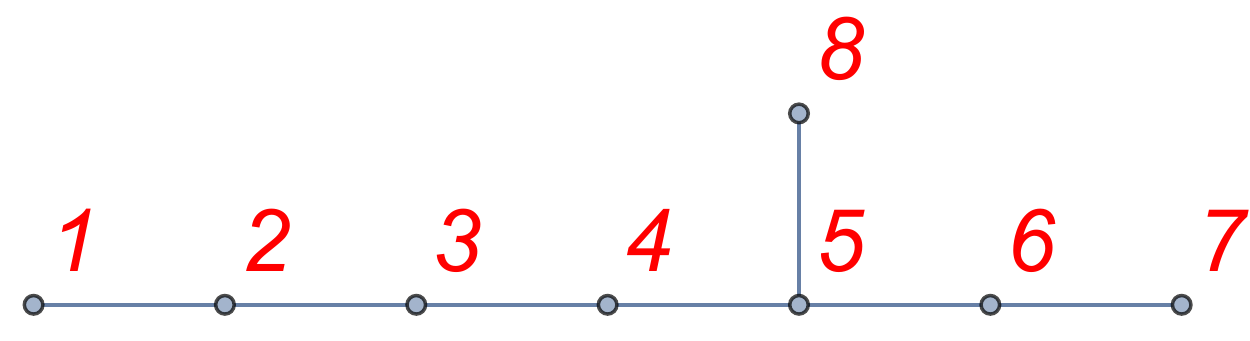}   $\mathring{\rho}=-\sum \bm{\alpha_j}+\rho = \{0, 1, 1, 1, 2, 1, 0, 0\}$.
} 

\smallskip

The decompositions of $R$ and $\hat R$ on the characters associated with the above weights are given in \cite{CZ1} for $A_r$, $r=1,2,3,4,5$, and in \cite{CMSZ} for $B_2, B_3$.
For $A_2$, aka $\SU(3)$, one may notice the remarkably simple result $R_3=\hat R_3 = 1$.
The result for $G_2$ (exercise for the reader!)  is $R=\hat R = 1/9 \, \chi_{0, 0} + 13/144 \, \chi_{0, 1} + 1/432 \, \chi_{0, 2} + 1/72 \, \chi_{1, 0}$.
The result for $A_6$, aka $\SU(7)$, is given in the present paper (eq.~\ref{R7result}).

\section*{Acknowledments}
The author wants to thank J.-B. Zuber for his comments on the manuscript.


 \end{document}